\documentclass[acp, manuscript]{copernicus}
\usepackage{graphicx,natbib,bm,url,color,colortbl}
\usepackage{breakurl}
\graphicspath{{./fig/}{./png/}}

\begin{document}

\title{Cloud-droplet growth due to supersaturation fluctuations in stratiform clouds}
\Author[1,2,3,4,5]{Xiang-Yu}{Li}
\affil[1]{Department of Meteorology and Bolin Centre for Climate Research, Stockholm University, Stockholm, Sweden}
\affil[2]{Nordita, KTH Royal Institute of Technology and Stockholm University, 10691 Stockholm, Sweden}
\affil[3]{Swedish e-Science Research Centre, Stockholm, Sweden}
\affil[4]{Laboratory for Atmospheric and Space Physics, University of Colorado, Boulder, CO 80303, USA}
\affil[5]{JILA, Box 440, University of Colorado, Boulder, CO 80303, USA}
\Author[1,3,6]{Gunilla}{Svensson}
\affil[6]{Global \& Climate Dynamics, National Center for Atmospheric Research, Boulder, CO 80305, USA}
\Author[2,4,5,7]{Axel}{Brandenburg}
\affil[7]{Department of Astronomy, Stockholm University, SE-10691 Stockholm, Sweden}
\Author[8,9]{Nils E. L.}{Haugen}
\affil[8]{SINTEF Energy Research, 7465 Trondheim, Norway}
\affil[9]{Department of Energy and Process Engineering, NTNU, 7491 Trondheim, Norway}

\runningtitle{Cloud droplet growth due to supersaturation fluctuations}

\runningauthor{Xiang-Yu Li}
\correspondence{Xiang-Yu Li (xiang.yu.li@su.se), \today,~ $ $Revision: 1.217 $ $}

\maketitle
\nolinenumbers
\begin{abstract}

Condensational growth of cloud droplets due to
supersaturation fluctuations
is investigated by solving the hydrodynamic and
thermodynamic equations using direct numerical simulations
with droplets being modeled as Lagrangian particles.
The supersaturation field is calculated directly by
simulating the temperature and water vapor fields instead of
being treated as a passive scalar.
Thermodynamic feedbacks to the fields due to condensation
are also included for completeness.
We find that the width of droplet size distributions increases with
time, which is contrary to the classical theory without
supersaturation fluctuations, where condensational
growth leads to progressively narrower size distributions.
Nevertheless, in agreement with earlier Lagrangian stochastic
models of the condensational growth, the standard deviation of the surface
area of droplets increases as $t^{1/2}$.
Also, for the first time, we explicitly demonstrate that the time evolution of the size distribution
is sensitive to the Reynolds number, but insensitive
to the mean energy dissipation rate.
This is shown to be due to the
fact that temperature fluctuations and water vapor mixing ratio fluctuations
increases with increasing Reynolds number, therefore the resulting
supersaturation fluctuations are enhanced with
increasing Reynolds number. Our simulations may explain the broadening
of the size distribution in stratiform clouds qualitatively, where the mean updraft
velocity is almost zero.

\end{abstract}

\section{Introduction}

The growth of cloud droplets is dominated by
two processes: condensation and collection.
Condensation of water vapor on active cloud condensation nuclei
is important in the size range from the activation size of aerosol particles to about
a radius of $10\,\mu{\rm m}$ \citep{pruppacher2012microphysics,2011_lamb}.
Since the rate of droplet growth by condensation
is inversely proportional to the droplet radius, large droplets grow slower than smaller ones.
This generates narrower size distributions \citep{2011_lamb}.
To form rain droplets in warm clouds, small droplets must grow to about $50\,\mu{\rm m}$ in radius
within 15--20 minutes \citep{pruppacher2012microphysics,DBB12,Grabowski_2013,Sei06}.
Therefore, collection, a widely accepted microscopical mechanism, has been
proposed to explain the rapid formation of rain droplets \citep{1955_Saffman, berry_1974, shaw_2003, Grabowski_2013}.
However, collection can only become active when the size distribution reaches a certain width. 

\cite{hudson1995cloud} observed a broadening of the droplet
size distribution in Californian marine stratus, which was contrary
to the classical theory of condensational growth \citep{yau1996short}.
The increasing width of droplet size distributions were further
observed by \cite{pawlowska2006observations} and \cite{Siebert17}.
The contradiction between the observed broadening width and the theoretical
narrowing width in the absence of turbulence has stimulated several studies.
The classical treatment of diffusion-limited growth assumes that supersaturation
depends only on average temperature and water mixing ratio.
Since fluctuations of temperature and the water mixing ratio are
affected by turbulence, the supersaturation fluctuations are inevitably subjected to
turbulence. Naturally, condensational growth due to supersaturation fluctuations
became the focus \citep{Sedunov65,kabanov1970effect,
cooper1989effects,srivastava1989growth,korolev1995influence,Khvorostyanov99,
2015_Sardina,grabowski2017broadening}.
The supersaturation fluctuations are particularly important for understanding
the condensational growth of cloud droplets in stratiform clouds,
where the updraft velocity of the parcel is almost zero
\citep{hudson1995cloud, korolev1995influence}.
When the mean updraft velocity is not zero, there could be a competition between
mean updraft velocity and supersaturation fluctuations.
This may diminish the role of supersaturation
fluctuations \citep{sardina2018broadening}.

Condensational growth due to supersaturation fluctuations 
was first recognized by \cite{srivastava1989growth}, who
criticized the use of a volume-averaged supersaturation
and proposed a randomly distributed supersaturation field.
\citet{cooper1989effects} proposed that droplets moving in clouds are exposed to
a varying supersaturation field. This results in broadening
of droplets size distribution due to supersaturation fluctuations.
\citet{Grabowski_2013} called the mechanism of \citet{cooper1989effects}
the eddy-hopping mechanism, which was then investigated by \citet{grabowski2017broadening}.
Using direct numerical simulations (DNS), \cite{vaillancourt2002microscopic}
found that the mean energy dissipation rate of turbulence has a negligible effect on condensational growth
and attributed this to the decorrelation between the supersaturation
and the droplet size.
\cite{paoli2009turbulent} considered three-dimensional (3-D)
turbulence as well as stochastically forced temperature and vapor
fields with a focus on statistical modeling for large-eddy simulations.
They found that supersaturation fluctuations due to turbulence mixing
are responsible for the broadening of the droplet size distribution.
\cite{2009_Lanotte} conducted 3-D DNS for condensational growth by only
solving a passive scalar equation for the supersaturation and concluded
that the width of the size distribution
increases with increasing Reynolds number. \cite{2015_Sardina} extended the DNS
of \cite{2009_Lanotte} to higher Reynolds number and found that the variance
of the size distribution increases in time. 
In a similar manner as
\cite{2015_Sardina}, \cite{siewert2017statistical} modelled the supersaturation field
as a passive scalar coupled to the Lagrangian particles and found that
their results can be reconciled with those of earlier numerical studies
by noting that the droplet size distribution broadens with increasing Reynolds number
\citep{paoli2009turbulent,2009_Lanotte, 2015_Sardina}.
Neither \cite{2015_Sardina} nor \cite{siewert2017statistical}
solved the thermodynamics that determine the supersaturation field. 
Both \cite{saito2017turbulence} and \cite{chen2018turbulence} solved
the thermodynamics equations governing the supersaturation field.
However, since collection was also included in their work,
one cannot clearly identify the roles of turbulence on collection
or condensational growth, nor can one compare their results with
Lagrangian stochastic models \citep{2015_Sardina, siewert2017statistical}
related to condensational growth.

Recent laboratory experiments and observations about cloud microphysics
also confirm the notion that supersaturation fluctuations may play
an important role in broadening the size distribution of cloud
droplets. The laboratory studies of \citet{Chandrakar16} and \citet{desai2018influence}
suggested that supersaturation fluctuations in the low aerosol number concentration limit
are likely of leading importance for the onset of precipitation.
The condensational growth due to supersaturation fluctuations seems to be
more sensitive to the integral scale of turbulence \citep{Gotzfried17}.
\cite{Siebert17} measured the variability of temperature, water vapor
mixing ratio, and supersaturation in warm clouds and support the notion
that both aerosol particle activation and droplet growth take place in the
presence of a broad distribution of supersaturation
\citep{hudson1995cloud,brenguier1998improvements,miles2000cloud,pawlowska2006observations}.
The challenge is now how to interpret the observed broadening
of droplet size distribution in warm clouds. How does turbulence drive
fluctuations of the scalar fields (temperature and water vapor mixing ratio)
and therefore affect the broadening of droplet size distributions 
\citep{Siebert17}?

In an attempt to answer this question, we conduct 3-D DNS experiments 
of condensational growth of cloud droplets, where turbulence, thermodynamics,
feedback from droplets to the fields via the condensation rate 
and buoyancy force are all included. The main aim is to investigate
how supersaturation fluctuations affect the droplet size
distribution.
We particularly focus on the time evolution of the size distribution $f(r,t)$
and its dependency on small and large scales of turbulence. We then compare our simulation results
with Lagrangian stochastic models \citep{2015_Sardina, siewert2017statistical}.
For the first time,
the stochastic model and simulation results from
the complete set of equations governing the supersaturation field are compared.

\section{Numerical model}

We now discuss the basic equations where we combine
the Eulerian description of the density ($\rho$),
turbulent velocity ($\bm{u}$), temperature ($T$),
and water vapor mixing ratio ($q_v$) with the Lagrangian
description of the ensemble of
cloud droplets.
The water vapor mixing ratio $q_v$ is defined as the ratio between the
mass density of water vapor and dry air.
Droplets are treated as superparticles. 
A superparticle represents an ensemble of droplets, whose mass, radius, and velocity
are the same as those of each individual droplet within it
\citep{Shima09, Johansen_2012, li2017eulerian}. For condensational growth, the
superparticle approach \citep{li2017eulerian} is the same as the Lagrangian
point-particle approach \citep{Kumar14} since there is no interactions among droplets.
Nevertheless, we still use the superparticle approach so that we can include
more processes like collection \citep{li2017eulerian, li2017effect}
in future.
Another reason to adopt superparticle approach is that it can be easily adapted
to conduct Large-eddy simulations with appropriate sub-grid scale models
\citep{grabowski2017broadening}.
To investigate the condensational growth of cloud droplets that
experience fluctuating supersaturation, 
we track each individual superparticle in a
Lagrangian manner. 
The motion of each superparticle is governed by the
momentum equation for inertial particles. 
The supersaturation
field in the simulation domain is determined by
$T(\bm{x},t)$ and $q_v(\bm{x},t)$ transported by turbulence. Lagrangian
droplets are exposed in different supersaturation fields. Therefore, droplets
either grow by condensation or shrink by evaporation depending on the local supersaturation field.
This phase transition generates a buoyancy force, which in turn
affects the turbulent kinetic energy, $T(\bm{x},t)$, and $q_v(\bm{x},t)$.
{\sc Pencil Code} \citep{axel_brandenburg_2018_2315093} is used to conduct all the simulations.

\subsection{Equations of motion for Eulerian fields}
\label{sec:model}

The background air flow is almost incompressible and thus obeys the Boussinesq approximation. 
Its density $\rho(\bm{x}, t)$ is governed by the continuity equation
and velocity $\bm{u}(\bm{x}, t)$ by Navier-Stokes equation.
The temperature $T(\bm{x}, t)$ of the background air flow is determined by the
energy equation with a source term due to the latent heat release.
The water vapor mixing ratio $q_v(\bm{x}, t)$ is transported by the background air flow.
The Eulerian equations are given by
\begin{equation}
\label{eq:continuity}
  {\partial\rho\over\partial t}+{\bm{\nabla}}\cdot(\rho\bm{u})=S_\rho ,
\end{equation}
\begin{equation}
{D\bm{u}\over D t}=\bm{f}
-\rho^{-1}{\bm{\nabla}} p
  +\rho^{-1} {\bm \nabla} \cdot (2 \nu \rho {\sf S})+B\bm{e}_z+{\bm S}_u ,
\label{turb}
\end{equation}
\begin{equation}
\label{eq:tt}
{D T\over Dt}
=\kappa\nabla^2 T+\frac{L}{c_p}C_d ,  
\end{equation}
\begin{equation}
{D q_v\over Dt}=D\nabla^2 q_v-C_d, 
\label{mixingRatio}
\end{equation}
where $D/Dt=\partial/\partial t+\bm{u}\cdot{\bm{\nabla}}$ is the material derivative,
${\bm f}$ is a random forcing function \citep{Haugen_etal_2004PhRvE},
$\nu$ is the kinematic viscosity of air,
${\sf S}_{ij}={\textstyle\frac{1}{2}}(\partial_j u_i+\partial_i u_j)
-{\textstyle{1\over3}}\delta_{ij}(\partial_k u_k)$ is
the traceless rate-of-strain tensor, $p$ is the gas pressure, $\rho$ is the gas density,
$c_{\rm p}$ is the specific heat at constant pressure, $L$ is
the latent heat, 
$\kappa$ is the thermal diffusivity of air,
$C_d$ is the condensation rate, $B$ is the buoyancy, $e_z$ is the
unit vector in the $z$ direction (vertical direction),
and $D$ is the diffusivity of water vapor.
To avoid global transpose operations associated with calculating
Fourier transforms for solving the nonlocal equation for the pressure in strictly
incompressible calculations, we solve here instead the compressible Navier-Stokes
equations using high-order finite differences.
The sound speed $c_{\rm s}$ obeys
$c_{\rm s}^2=\gamma p/\rho$, where $\gamma=c_{\rm p}/c_{\rm v}=7/5$
is the ratio between specific heats, $c_{\rm p}$ and $c_{\rm v}$, at constant
pressure and constant volume, respectively.
We set the sound speed as $5\,\rm{m} \, \rm{s}^{-1}$
to simulate the nearly incompressible atmospheric air flow,
resulting in a Mach number
of $0.06$ when $u_{\rm rms}=0.27\,\rm{m} \, \rm{s}^{-1}$, where $u_{\rm rms}$
is the rms velocity.
Such a configuration, with so small Mach number, is almost equivalent 
to an incompressible flow.
It is worth noting that the temperature determining the compressibility of
the flow is constant and independent of the temperature field of the gas
flow governed by Equation~(\ref{eq:tt}).
Also, since the gas flow is almost incompressible and its mass density
is much smaller than the one of the droplet, there is no mass
exchange between the gas flow and the droplet, i.e.,
the density of the gas flow $\rho(\bm{x},t)$ is not affected by
$T(\bm{x},t)$.
Thus, the source terms $S_\rho$ and $\bm{S}_u$ in
Equations~(\ref{eq:continuity}) and (\ref{turb})
are neglected \citep{kruger2017correlation}.
The buoyancy $B(\bm{x},t)$ depends on the temperature $T(\bm{x},t)$,
water vapor mixing ratio $q_v(\bm{x},t)$,
and the liquid mixing ratio $q_l$ \citep{Kumar14}, 
\begin{equation}
B(\bm{x},t)=g(T^{\prime}/T+\alpha q_v^{\prime}-q_l),
\end{equation}
where $\alpha=M_a/M_v-1\approx0.608$ when $M_a$
and $M_v$ are the molar masses of air and water vapor, respectively.
The amplitude of the gravitational acceleration is given by $g$.
The liquid water mixing ratio is the ratio between the mass density
of liquid water and the dry air and is defined as 
\begin{equation}
q_l\left(\bm{x},t\right)= 
\frac{4\pi\rho_l}{3\rho_a (\Delta x)^3}\sum_{j=1}^{N_{\triangle}}  
r\left(t\right)^3
=\frac{4\pi\rho_l}{3\rho_a }\sum_{j=1}^{N_{\triangle}}  
f(r,t)r\left(t\right)^3\delta r
, 
\end{equation}
where $\rho_l$ and $\rho_a$ are the liquid water density 
and the reference mass density of dry air.
$N_{\triangle}$ is the total 
number of droplets in a cubic grid cell with volume $(\Delta x)^3$,
where $\Delta x$ is the one-dimensional size of the grid box.
The temperature fluctuations are given by
\begin{equation}
T^{\prime}\left(\bm{x},t\right)=T\left(\bm{x},t\right)-T_{\rm env},
\end{equation}
and the water vapor mixing ratio fluctuations by
\begin{equation}
q_v^{\prime}\left(\bm{x},t\right)=q_v\left(\bm{x},t\right)-q_{v,\rm{env}}.
\end{equation}
We adopt the same method as in \cite{Kumar14},
where the mean environmental temperature $T_{\rm env}$ and water vapor mixing ratio $q_{v,\rm{env}}$
do not change in time. This assumption is plausible in the circumstance that we
do not consider the entrainment, i.e., there is only mass and energy
transfer between liquid water and water vapor.
The condensation rate $C_d$ \citep{vaillancourt2001microscopic} is given by
\begin{equation}
C_d\left(\bm{x},t\right)=
\frac{4\pi\rho_l G}{\rho_a (\Delta x)^3}\sum_{j=1}^{N_{\triangle}}  
s\left(\bm{x},t\right)r\left(t\right)= 
\frac{4\pi\rho_l G}{\rho_a }\sum_{j=1}^{N_{\triangle}}
s\left(\bm{x},t\right)f(\bm{x},t)r\left(t\right)\delta r,
\label{condensationRate}
\end{equation}
where $G$ is the condensation parameter
(in units of ${\rm m}^2\,{\rm s}^{-1}$), which depends weakly on temperature and pressure
and is here assumed to be constant \citep{2011_lamb}.
The supersaturation $s$ is defined as
the ratio between the vapor pressure $e_v$ and the saturation vapor
pressure $e_s$,
\begin{equation}
\label{eq: saturation}
s = \frac{e_v}{e_s}-1.
\end{equation}
Using the ideal gas law, Equation~(\ref{eq: saturation}) can be expressed
as,
\begin{equation}
\label{eq: saturation2}
s = \frac{\rho_v R_v T}{\rho_{vs} R_v T}-1=\frac{\rho_v}{\rho_{vs}}-1.
\end{equation}
In terms of the water vapor mixing ratio $q_v=\rho_v/\rho_a$ and saturation water
vapor mixing ratio $q_{vs}=\rho_{vs}/\rho_a$, Equation~(\ref{eq: saturation2}) can
be written as:
\begin{equation}
s\left(\bm{x},t\right)=\frac{q_v\left(\bm{x},t\right)}{q_{vs}\left(T\right)}-1. 
\label{supersaturation}
\end{equation}
Here $\rho_v$ is the mass density
of water vapor and $\rho_{vs}$ the mass density of saturated water vapor,
and $q_{vs}\left(T\right)$ is the
saturation water vapor mixing ratio at temperature $T$ and
can be determined by the ideal gas law,
\begin{equation}
q_{vs}\left(T\right)=\frac{e_s\left(T\right)}{R_v \rho_a T}. 
\label{eq: qvs}
\end{equation}
The saturation vapor pressure $e_s$ over liquid water
is the partial pressure due to the water vapor when an
equilibrium state of evaporation and condensation is reached
for a given temperature.
It can be determined by the Clausius-Clapeyron equation, which
determines the change of $e_s$ with temperature $T$.
Assuming constant latent heat $L$, $e_s$ is approximated as \citep{yau1996short, Gotzfried17}
\begin{equation}
\label{eq:e_s}
e_s(T)=c_1 \exp(-c_2/T),
\end{equation}
where $c_1$ and $c_2$ are constants
adopted from page~14 of \citet{yau1996short}.
We refer to Table~\ref{constants} for all the thermodynamics constants.
In the present study, the updraft cooling is omitted.
Therefore, the assumption of constant latent heat $L$ is plausible.

\subsection{Lagrangian model for cloud droplets}

In addition to the Eulerian fields described in Section~\ref{sec:model}
we treat cloud droplets as Lagrangian particles.
In the {\sc Pencil Code}, they are invoked as non-interacting superparticles.

\subsubsection{Kinetics of cloud droplets}

Each superparticle is treated as a
Lagrangian point-particle, where one solves for the particle position
$\bm{x}_i$,
\begin{equation}
  \frac{d\bm{x}_i}{dt}=\bm{V}_i,
\end{equation}
and its velocity $\bm{V}_i$ via
\begin{equation}
  \frac{d\bm{V}_i}{dt}=\frac{1}{\tau_i}(\bm{u}-\bm{V}_i)+g\bm{e}_z,
\end{equation}
in the usual way; see \citep{li2017eulerian} for details.
Here, $\bm{u}$ is the fluid velocity at the position of the superparticle,
$\tau_i$ is the particle inertial response or stopping time
of a droplet $i$ and is given by
\begin{equation}
\label{response_time}
\tau_i=2\rho_{\rm l} r_i^2/[9\rho\nu \, D({\rm Re}_i)].
\end{equation}
The correction factor \citep{Schiller33, Marchioli08},
\begin{equation}
D({\rm Re}_i)=1+0.15\,{\rm Re}_i^{2/3},
\label{eq:correction}
\end{equation}
models the effect of non-zero particle Reynolds number
${\rm Re}_i=2r_i|\bm{u}-\bm{V}_i|/\nu$.
This is a widely used approximation, although it does not
correctly reproduce the small-$\rm{Re}_i$ correction to Stokes formula \citep{John07}.

\subsubsection{Condensational growth of cloud droplets}

The condensational growth of the particle radius $r_i$
is governed by \citep{pruppacher2012microphysics,2011_lamb}
\begin{equation}
	{{\rm d} r_i\over{\rm d} t}={Gs\left(\bm{x}_i,t\right)\over r_i}.
\label{cond_eq}
\end{equation}

\section{Experimental setup}

\subsection{Initial configurations}

The initial values of the water vapor mixing ratio
$q_v(\bm{x}, t=0)=0.0157 \,{\rm kg} {\rm kg}^{-1}$ and temperature $T(\bm{x}, t=0)=292\,{\rm K}$
are matched to the ones obtained in the CARRIBA experiments \citep{katzwinkel2014measurements},
which are the same as those in \cite{Gotzfried17}.
With this configuration, we obtain $s(\bm{x}, t=0)=2\%$, which means that the
water vapor is initially supersaturated.
The time step of the simulations presented here is governed by the smallest time scale
in the present configuration, which is the particle stopping time defined
in Equation~(\ref{response_time}).
The thermodynamic time scale is much larger than the turbulent one.
Table~\ref{constants} shows the list of thermodynamic parameters used in the present study.

Initially, 10 $\mu$m-sized droplets with zero velocity
are randomly distributed in the simulation domain.
The mean number density of droplets, which is constant
in time since droplet collections are not considered, 
is $n_0=2.5\times10^8\rm{m}^{-3}$.
This gives an initial liquid water content,
$\int_{0}^\infty f(r,t=0)\,r^3 \,{\rm d}r$,
which is $0.001\,{\rm kg\, m}^{-3}$.
The simulation domain is a cube of size $L_x=L_y=L_z$, the values of which
are given in Table~\ref{Swarm_Rey}.
The number of superparticles $N_{\rm s}$ satisfies
$N_{\rm s}/N_{\rm grid}\approx0.1$, where $N_{\rm grid}$
is the number of lattices depending on the spatial resolution of the simulations. 
Setting $N_{\rm s}/N_{\rm grid}\approx0.1$, on one hand, is still within the convergence
range $N_{\rm s}/N_{\rm grid}\approx0.05$ \citep{li2017effect}.
On the other hand, it can mimic the diluteness of the atmospheric cloud system,
where there are about 0.1 droplets per cubic Kolmogorov scale.
This configuration results in $N_{\rm s, 128}=244140$ when $N_{\rm grid}=128^3$.

\begin{table}[t!]
\caption{List of constants for the thermodynamics: see text for explanations of symbols.}
\centering
\setlength{\tabcolsep}{3pt}
\begin{tabular}{lr}
Quantity & Value \\
\hline
  $\nu$ ($\rm{m}^2\rm{s}^{-1}$) & $1.5\times10^{-5}$ \\ 
  $\kappa$ ($\rm{m}^2\rm{s}^{-1}$) & $1.5\times10^{-5}$ \\ 
$D$ ($\rm{m}^2\rm{s}^{-1}$) & $2.55\times10^{-5}$ \\ 
$G$ ($\rm{m}^2\rm{s}^{-1}$) & $1.17\times10^{-10}$ \\ 
$c_1$ (Pa) & $2.53\times10^{11}$ \\ 
$c_2$ (${\rm K}$) & 5420 \\ 
$L$ (${\rm J} {\rm kg}^{-1}$) & $2.5\times10^6$ \\ 
$c_{\rm p}$ (${\rm J} {\rm kg}^{-1} {\rm K}^{-1}$) & 1005.0 \\ 
$R_v$ (${\rm J} {\rm kg}^{-1} {\rm K}^{-1}$) & 461.5 \\ 
$M_a$ (${\rm g} {\rm mol}^{-1}$) & 28.97 \\
$M_v$ (${\rm g} {\rm mol}^{-1}$) & 18.02 \\
$\rho_a$ (${\rm kg} {\rm m}^{-3}$) & 1 \\
$\rho_l$ (${\rm kg} {\rm m}^{-3}$) & 1000 \\
$\alpha$ & 0.608 \\
${\rm Pr}=\nu/\kappa$ & 1 \\
$Sc=\nu/D$ & 0.6 \\
$q_v(\bm{x}, t=0)$ ($\,{\rm kg} {\rm kg}^{-1}$) & 0.0157 \\
$q_{v,\rm{env}}\,({\rm kg} {\rm kg}^{-1})$ & 0.01\\
$T(\bm{x}, t=0)\,({\rm K})$ & 292 \\
$T_{\rm{env}}\,({\rm K})$ & 293 \\ 
\hline
\end{tabular}
\label{constants}
\end{table}

\subsection{DNS}

We conduct high resolution simulations \citep{li_xiang_yu_2019_2538027} for different
Taylor micro-scale Reynolds number ${\rm Re}_{\lambda}$
and mean energy dissipation rate $\bar{\epsilon}$
(see Table~\ref{Swarm_Rey} for details of the simulations).
The Taylor micro-scale Reynolds
number is defined as ${\rm Re}_\lambda \equiv u_{\rm rms}^2 \sqrt{5/(3\nu\bar{\epsilon})}$.
For simulations with different values of $\bar{\epsilon}$ at fixed ${\rm Re}_{\lambda}$,
we vary both the domain size $L_x$ ($L_y=L_z=L_x$) and the amplitude of the forcing $f_0$.
As for fixed $\bar{\epsilon}$, ${\rm Re}_{\lambda}$ is varied by solely changing
the domain size, which in turn changes $u_{\rm rms}$.
In all simulations, we use for the Prandtl number ${\rm Pr}=\nu/\kappa=1$
and for the Schmidt number $Sc=\nu/D=0.6$.
For our simulations with $N_{\rm grid}=512^3$ meshpoints,
the code computes 55,000 time steps in 24 hours wall-clock time using 4096 cores.
For $N_{\rm grid}=128^3$ meshpoints, the code computes 4.5 million
time steps in 24 hours wall-clock time using 512 cores.

\section{Results}

\begin{table*}[t!]
\caption{Summary of the simulations; see text for explanation of symbols.}
\centering
\setlength{\tabcolsep}{3pt}
\begin{tabular}{lcccccccccccr}
  Run &  $f_0$ & $L_x\,(\rm{m})$ & $N_{\rm grid}$ & $N_{\rm s}$& $u_{\rm rms}$ ($\rm{m} \, \rm{s}^{-1}$) & $\mbox{\rm Re}_{\lambda}$ & $\bar{\epsilon}$ ($\rm{m}^2\rm{s}^{-3}$)& $\eta$ ($10^{-4} \rm{m}$) & $\tau_{\eta}$ ($\rm{s}$) & $\tau_{\rm L}$ ($\rm{s}$) & $\tau_{\rm s}$ ($\rm{s}$) & $\rm{Da}$ \\
\hline
A & $0.02$  & 0.125 & $128^3$ & $N_{\rm s, 128}$ &0.16 &45  & 0.039 &5.4  & 0.020 & 0.25  & 0.014 & 0.053 \\ 
  B & $0.02$  & 0.25  & $256^3$ & $2^3N_{\rm s, 128}$&0.22 &78  & 0.039  & 5.4  & 0.020 & 0.37 & 0.014 & 0.081 \\ 
C & $0.02$  & 0.5  & $512^3$ & $2^6N_{\rm s, 128}$&0.28 &130  & 0.039  & 5.4  & 0.020 & 0.58 & 0.014 & 0.125 \\ 
D & $0.014$  & 0.6 & $512^3$ & $2^6N_{\rm s, 128}$ &0.24 &135  & 0.019 & 6.5  & 0.028 & 0.81 & 0.014 & 0.174 \\ 
E & $0.007$  & 0.8 & $512^3$ & $2^6N_{\rm s, 128}$ &0.17 &138  & 0.005 &8.9  & 0.053 & 1.47  & 0.014 & 0.312 \\ 
\hline
\end{tabular}
\label{Swarm_Rey}
\end{table*}

\begin{figure*}[t!]
\begin{center}
\includegraphics[width=\textwidth]{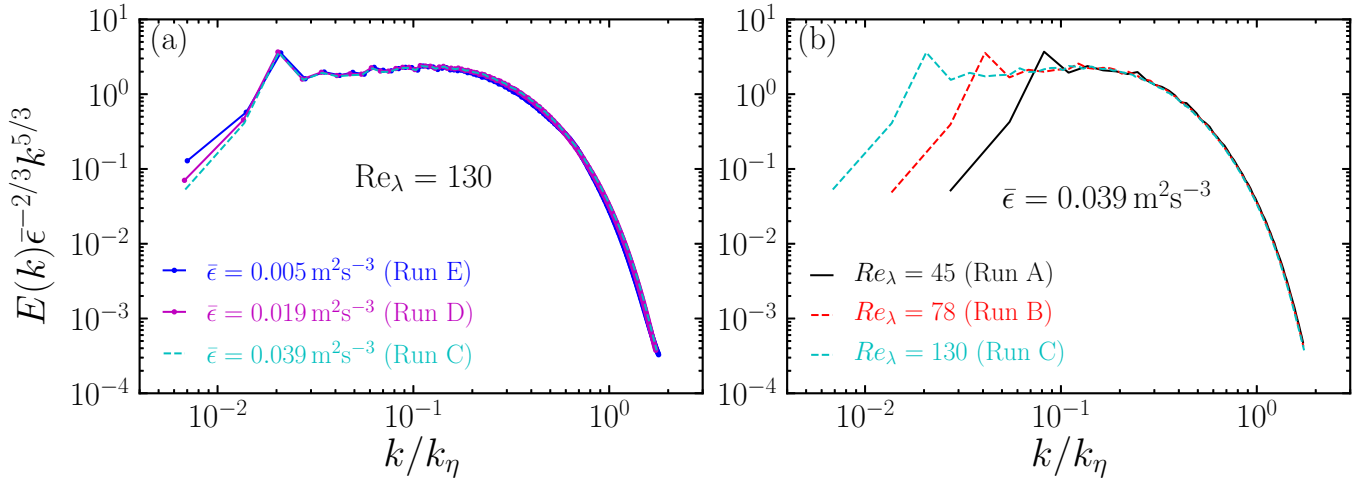}
\end{center}
\caption{Time-averaged kinetic energy spectra of the turbulence gas flow
for (a) different $\bar{\epsilon}$ = $0.005\,\rm{m}^2\rm{s}^{-3}$ (blue dash-dotted line),
0.019 (magenta dash-dotted line) and 0.039
(dashed cyan line) at fixed $\mbox{\rm Re}_{\lambda}=130$
(see Runs~C, D, and E in Table~\ref{Swarm_Rey} for details)
and for (b) different $\mbox{\rm Re}_{\lambda}$ = 45 (black solid line),
78 (red dashed line), and 130 (cyan dashed line) at fixed
$\bar{\epsilon}$ = $0.039\rm{m}^2\rm{s}^{-3}$
(see Runs~A, B, and C in Table~\ref{Swarm_Rey} for details).
}
\label{ppower_FixE0p03}
\end{figure*}

Figure~\ref{ppower_FixE0p03}(a) shows time-averaged turbulent kinetic-energy spectra
for different values of $\bar{\epsilon}$ at fixed ${\rm Re}_{\lambda}\approx130$.
Since the abscissa in the figures is normalized by $k_\eta=2 \pi /\eta$,
the different spectra shown in Figure~\ref{ppower_FixE0p03}(a) collapse onto a single curve.
Here, $\eta$ is the Kolmogorov length scale.
Figure~\ref{ppower_FixE0p03}(b) shows
the time-averaged turbulent kinetic-energy spectra
for different values of ${\rm Re}_{\lambda}$ at fixed
$\bar{\epsilon}\approx0.039\,\rm{m}^2\rm{s}^{-3}$.
For larger Reynolds numbers the spectra extend to smaller wavenumbers.
A flat profile corresponds to Kolmogorov scaling \citep{Pope00}
when the energy spectrum is compensated by $\bar{\epsilon}^{-2/3}k^{5/3}$.
For the largest ${\rm Re}_{\lambda}$ in our simulations
($\mbox{\rm Re}_\lambda=130$),
the inertial range extends for about a decade
in $k$-space.

\begin{figure}[t!]\begin{center}
\includegraphics[width=\textwidth]{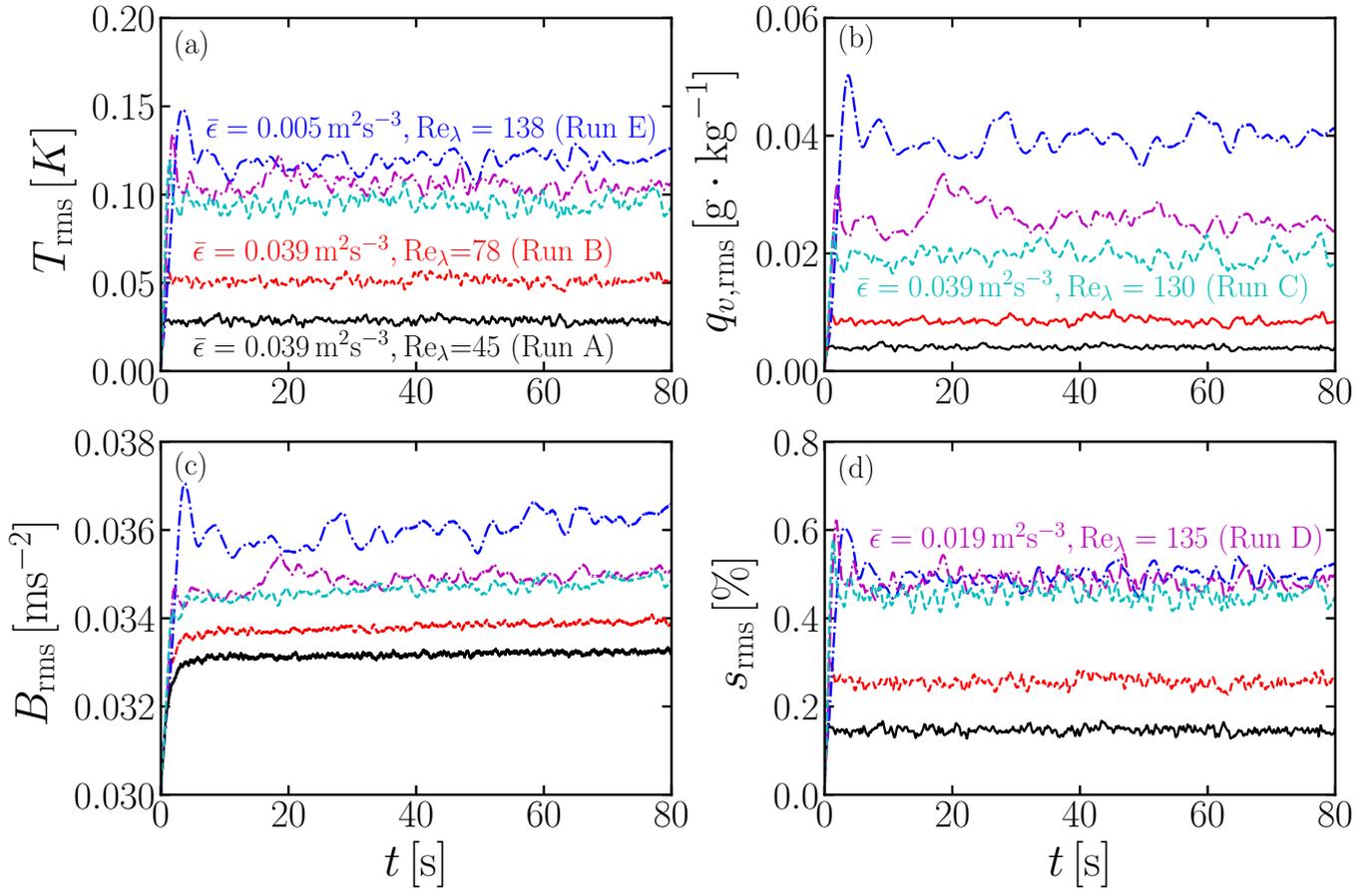}
\end{center}\caption[]{Time series of the field quantities:
(a) $T_{\rm rms}$, (b) $q_{v},\rm{rms}$, (c) $B_{\rm rms}$, and (d) $s_{\rm rms}$.
Same simulations as in Figure~\ref{ppower_FixE0p03}.
} 
\label{timeSeries_gravRe50_cond_e_comp}
\end{figure}

Next we inspect the response of thermodynamics to turbulence.
In Figure~\ref{timeSeries_gravRe50_cond_e_comp}, we show time series of 
fluctuations of temperature $T_{\rm rms}$,
water vapor mixing ratio $q_{v,\rm rms}$,
buoyancy force $B_{\rm rms}$,
and the supersaturation $s_{\rm rms}$.
All quantities reach a statistically steady state within a few seconds.
The steady state values of $T_{\rm rms}$, $q_{v,\rm rms}$, and
$s_{\rm rms}$ increase with increasing ${\rm Re}_{\lambda}$
approximately linearly, and vary hardly at all with $\bar{\epsilon}$.
On the other hand. $B_{\rm rms}$ changes only by a few percent as
${\rm Re}_{\lambda}$ or $\bar{\epsilon}$ vary.
Note, however, that the buoyancy force is only about 0.3\% of the
fluid acceleration.
This is because $T_{\rm rms}$ is small (about $0.1\,\rm{K}$
in the present study). Therefore, the effect of the buoyancy force should
indeed be small.

When changing $\bar{\epsilon}$ while keeping ${\rm Re}_{\lambda}$ fixed, the Kolmogorov scales
of turbulence varies.
Therefore, the various fluctuations quoted above are insensitive to the
small scales of turbulence.
However, when varying ${\rm Re}_{\lambda}$ while keeping $\bar{\epsilon}$ fixed,
their rms values change, which is due to large scales of turbulence.
Indeed, temperature fluctuations are driven by the large scales of turbulence,
which affects the supersaturated vapor pressure $q_{vs}$ via the
Clausius-Clapeyron equation; see Equation~(\ref{eq: qvs}).
Therefore, supersaturation fluctuations result from both temperature fluctuations
and water vapor fluctuations via Equation~(\ref{supersaturation}).
Both $q_{v,\rm{rms}}$ and $T_{\rm rms}$ increase with increasing ${\rm Re}_{\lambda}$,
resulting in larger fluctuations of $s$.
Supersaturation fluctuations, in turn, affect $T$ and $q_v$ via the
condensation rate $C_d$.

\begin{figure}[t!]\begin{center}
\includegraphics[width=\textwidth]{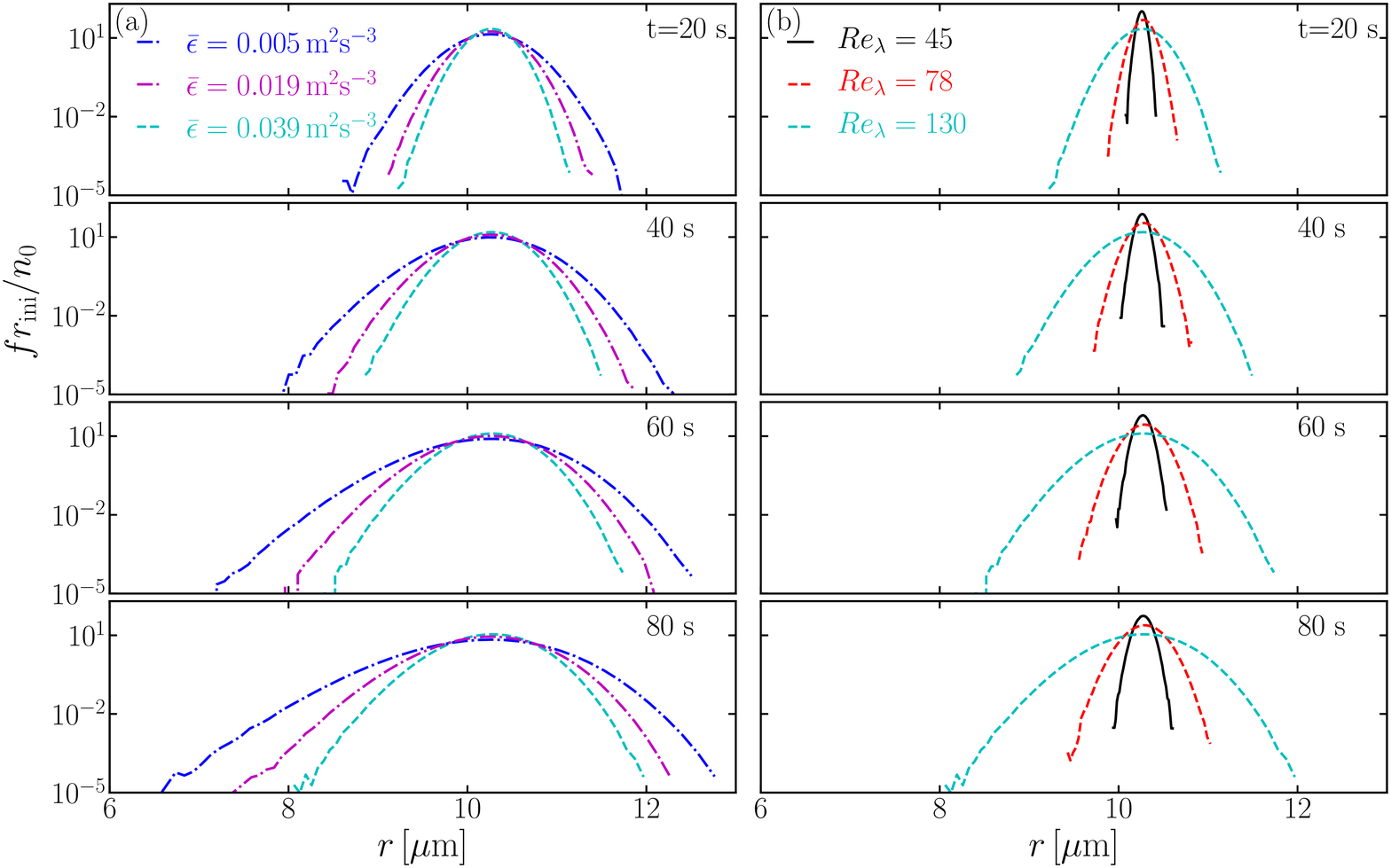}
\end{center}\caption[]{Comparison of the time evolution of droplet
size distributions for
different (a) $\bar{\epsilon}$ at $\rm{Re}_\lambda=130$ (Runs C, D, and E in Table~\ref{Swarm_Rey}) and
(b) ${\rm Re}_{\lambda}$ at $\bar{\epsilon}$ = $0.039\rm{m}^2\rm{s}^{-3}$ (Runs A, B, and C in Table~\ref{Swarm_Rey}).
Same simulations as in Figure~\ref{ppower_FixE0p03}.
} 
\label{f_gravRe50_cond_e_comp}
\end{figure}

Our goal is to investigate the condensational growth of cloud
droplets due to supersaturation fluctuations. 
Figure~\ref{f_gravRe50_cond_e_comp} shows the time evolution of droplet size
distributions for different configurations.
The conventional
understanding is that condensational growth leads to a narrow size distribution
\citep{pruppacher2012microphysics,2011_lamb}.
However, supersaturation fluctuations broaden the distribution.
More importantly, the width of the size
distribution increases with increasing ${\rm Re}_{\lambda}$,
but {\em decreases} slightly with increasing $\bar{\epsilon}$
over the range studied here. This is consistent with
the results shown in Figure~\ref{timeSeries_gravRe50_cond_e_comp}
in that supersaturation fluctuations are sensitive to
${\rm Re}_{\lambda}$ but are insensitive to $\bar{\epsilon}$.
In atmospheric clouds, ${\rm Re}_{\lambda}\approx10^4$,
which may result in an even broader size distribution.

\begin{figure}[t!]\begin{center}
\includegraphics[width=\textwidth]{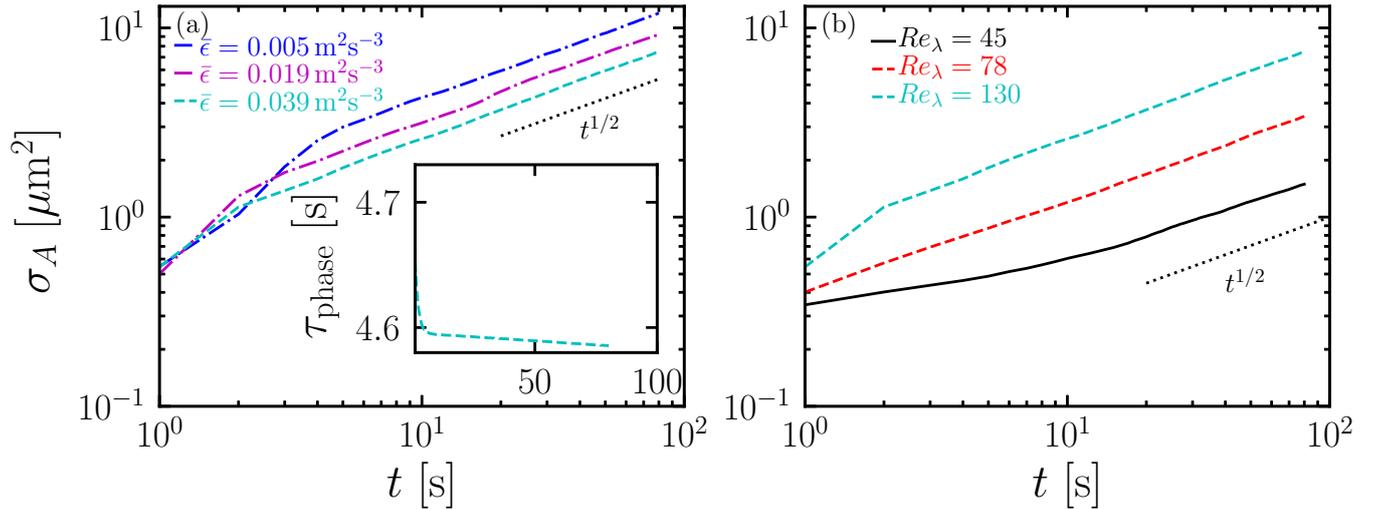}
\end{center}\caption[]{Time evolution of $\sigma_A$ for different
(a) $\bar{\epsilon}$ at $\rm{Re}_\lambda=130$ and
(b) ${\rm Re}_{\lambda}$ at $\bar{\epsilon}$ = $0.039\,\rm{m}^2\rm{s}^{-3}$.
Same simulations as in Figure~\ref{ppower_FixE0p03}.
} 
\label{sigma_area_f_comp}
\end{figure}

We further quantify the variance of the size distribution
by investigating the time evolution of the standard deviation of the droplet surface area
$\sigma_A$ for different configurations.
In terms of the droplet surface area $A_i$ ($A_i \propto r_i^2$),
Equation~(\ref{cond_eq})
can be written as
\begin{equation}
\frac{dA_i}{dt}=2Gs.
\label{eq:A}
\end{equation}
It can be seen from Equation~(\ref{eq:A}) that the evolution of the surface area
is analogous to Brownian motion,
indicating that its standard deviation $\sigma_A\propto\sqrt{t}$.
A more detailed stochastic model for $\sigma_A$ is developed by
\cite{2015_Sardina}.
Based on Equation~(\ref{cond_eq}), $\sigma_A$ is given by
\begin{equation}
\frac{d\sigma_A^2}{dt}=\frac{d}{dt}\left<A^{\prime 2}\right>
=\frac{d}{dt}\left<A^2-\left<A\right>^2\right>=4G\left<s^{\prime}A^\prime\right>.
\label{eq:area}
\end{equation}
\citet{2015_Sardina} adopted a Langevin equation to model the supersaturation field
and the vertical velocity of droplets, resulting in the scaling law:
\begin{equation}
\sigma_A \sim C(\tau_L, \tau_{\rm s}, {\rm Re}_\lambda)t^{1/2},
\label{eq:area2}
\end{equation}
where $C(\tau_L, \tau_{\rm s, {\rm Re}_\lambda})$ is a constant
for given $\tau_L$, $\tau_{\rm s}$, and ${\rm Re}_\lambda$.
Under the assumptions that $\tau_{\rm s}\ll T_L$
and a negligible influence on the macroscopic observables from small-scale
turbulent motions, \cite{2015_Sardina} obtained an analytical
expression for $\sigma_A$ as:
\begin{equation}
\sigma_A \sim \tau_{\rm s}{\rm Re}_{\lambda}t^{1/2},
\label{eq:area3}
\end{equation}
where $\tau_{\rm s}$ is the phase transition time scale given by
\begin{equation}
\tau_{\rm s}^{-1}(t) = 4 \pi G \int_{0}^{\infty} r f dr ,
\label{eq:phase}
\end{equation}
and $\tau_L$ is the turbulence integral time scale.
The model proposed that condensational growth of cloud droplets
depends only on ${\rm Re}_{\lambda}$ and is independent of $\bar{\epsilon}$.
In terms of the size distribution $f(r,t)$, $\sigma_A$ can be given as:
\begin{equation}
\sigma_A =\sqrt{a_4-a_2^2},
\label{eq:area4}
\end{equation}
where $a_{\zeta}$ is the moment of the size distribution,
which is defined as:
\begin{equation}
a_\zeta=\int_{0}^\infty f\,r^\zeta \,{\rm d}r\bigg/\int_{0}^\infty f\,{\rm d}r .
\label{azeta}
\end{equation}
Here, $\zeta$ is a positive integer.
As shown in Figure~\ref{sigma_area_f_comp}, the time evolution of $\sigma_A$ agrees
with the prediction $\sigma_A \propto t^{1/2}$.
\cite{2015_Sardina} and \cite{siewert2017statistical} solved the
passive scalar equation of $s$ without considering fluctuations
of $T$ and $q_v$. Feedbacks to flow fields from cloud droplets
were also neglected. They found good agreement between the DNS
and the stochastic model.
Comparing with \cite{2015_Sardina} and \cite{siewert2017statistical}, our study
solve the complete sets of the thermodynamics of supersaturation.
It is remarkable that a good agreement between the stochastic model
and our DNS is observed. This indicates that the stochastic model
is robust.
On the other hand, modeling supersaturation fluctuations
using the passive scalar equation seems to be sufficient
for the Reynolds numbers considered in this study.
We recall that $\tau_{\rm s}$ in Equation~(\ref{eq:area3}) is constant.
In the present study, $\tau_{\rm s}$ is determined by Equation~(\ref{eq:phase}).
Therefore, $\tau_{\rm s}$ varies with time as shown in the inset
of Figure~\ref{sigma_area_f_comp}(a). Nevertheless, since the variation of
$\tau_{\rm s}$ is small, we still observe $\sigma_A \sim t^{1/2}$
except for the initial phase of the evolution, where $s(t=0)=2\%$.

Comparing panels (a) and (b) of Figure~\ref{sigma_area_f_comp}, it is clear
that changing $\rm{Re}_\lambda$ has a much larger effect on $\sigma_A$
than changing $\bar{\epsilon}$.
In fact, as $\bar{\epsilon}$ is increased by a factor of about 8,
$\sigma_A$ decreases only by a factor of about 1.6, so the ratio of
their logarithms is about 1/5, i.e., $\sigma_A\propto\bar{\epsilon}^{\,-1/5}$.
By contrast, $\sigma_A$ changes by a factor of about 5 as $\rm{Re}_\lambda$ is
increased by a factor of nearly 3, so $\sigma_A\propto\rm{Re}_\lambda^{3/2}$.
This quantifies the high sensitivity of $\sigma_A$ to changes of
$\rm{Re}_\lambda$ compared to $\bar{\epsilon}$.

Two comments are here in order.
First, we emphasize that we observe here $\sigma_A\propto\rm{Re}_\lambda^{3/2}$
instead of $\sigma_A\propto\rm{Re}_\lambda$.
Therefore, there could be a critical $\rm{Re}_\lambda$,
beyond which $\sigma_A\propto\rm{Re}_\lambda$ and below which
$\sigma_A\propto\rm{Re}_\lambda^{3/2}$.
However, the highest $\rm{Re}_\lambda$ in our DNS is 130.
To verify this proposal, a large parameter range of $\rm{Re}_\lambda$ is required.
Second, we note that $\sigma_A\propto\bar{\epsilon}^{\,-1/5}$.
This is because the Damk\"{o}hler number increases with decreasing
$\bar{\epsilon}$ (see Table~\ref{Swarm_Rey}), which is defined as the ratio of
the fluid time scale to the characteristic thermodynamic time scale
associated with the evaporation process ${\rm Da}=\tau_{\rm L}/\tau_{\rm s}$.
\citet{vaillancourt2002microscopic} also found that $\sigma_A$ decreases
with $\bar{\epsilon}$, even though the mean updraft cooling is included
in their study.

\section{Discussion and conclusion}

Condensational growth of cloud droplets due to supersaturation fluctuations
is investigated using DNS. Cloud droplets are tracked in a 
Lagrangian framework,
where the momentum equation 
for inertial particles are solved. The thermodynamic
equations governing the supersaturation field are solved simultaneously.
Feedback from cloud droplets onto $\bm{u}$, $T$, and $q_v$ is included through
the condensation rate and buoyancy force.
We resolve the smallest scale of turbulence in all simulations.
Contrary to the classical condensation theory,
which leads to a narrow distribution when supersaturation fluctuations are
ignored, we find that droplet size distributions broaden due to
supersaturation fluctuations.
For the first time, we explicitly demonstrate that the size distribution becomes wider
with increasing ${\rm Re}_{\lambda}$, which is, however, insensitive to $\bar{\epsilon}$.
Supersaturation fluctuations are subjected to both temperature fluctuations
and water vapor mixing ratio fluctuations.

We observe that $\sigma_A \propto \sqrt{t}$ when the
complete sets of the thermodynamics equations governing the supersaturation are solved,
which are consistent with the findings by \cite{2015_Sardina} and \cite{siewert2017statistical}
even though fluctuations of temperature and water vapor mixing ratio, buoyancy force, and
droplets feedbacks to the field quantities
are neglected in their studies.
This indicates that the stochastic model of condensational
growth developed by \cite{2015_Sardina} is robust.
For the first time, to our knowledge,
the stochastic model \citep{2015_Sardina} and simulation results from
the complete set of thermodynamics equations governing the supersaturation
field are compared.
The broadening size distribution with increasing ${\rm Re}_{\lambda}$ demonstrates that condensational
growth due to supersaturation fluctuations is an important mechanism for droplet growth.
The maximum ${\rm Re}_{\lambda}$ in the present study is 130, which is about two orders
of magnitude smaller than the one in atmospheric clouds (${\rm Re}_{\lambda}=10^4$).
Since the width of the size distribution increases dramatically with increasing
${\rm Re}_{\lambda}$, the supersaturation fluctuation facilitated condensation may easily
overcome the bottleneck barrier \citep{Grabowski_2013}. 

The stochastic model developed by \cite{2015_Sardina} assumes that the width
of droplet size distributions is independent of $\bar{\epsilon}$.
Our result shows that the width decreases slightly with increasing $\bar{\epsilon}$.
However, the largest $\bar{\epsilon}$ in warm clouds is
about $10^{-3}\,\rm{m}^2\rm{s}^{-3}$ \citep{Grabowski_2013}. Therefore, neglecting
the smallest scales in the stochastic model is indeed acceptable.
\cite{vaillancourt2002microscopic} also found that the width of the droplet size
distribution decreases with increasing $\bar{\epsilon}$ and attributed this
to the decorrelation between supersaturation fluctuations and surface area
of droplets. \citet{2015_Sardina}, however, found stronger correlation
between supersaturation fluctuations and surface area
of droplets with increasing $\rm{Re}_\lambda$.
The present study is consistent with both the works of
\citet{vaillancourt2002microscopic} and \citet{2015_Sardina}.
Therefore, we emphasize that there is {\em{no}} contradiction between
both papers.

In the present study, the simulation box is stationary,
which means that the volume is not exposed to cooling, as no mean updraft is considered. 
Therefore, the condensational growth is
solely driven by supersaturation fluctuations. This is similar to the condensational
growth of cloud droplets in stratiform clouds, where the updraft velocity of
the parcel is close to zero \citep{hudson1995cloud, korolev1995influence}.
The observational data shows that the width of the size distribution is wider than the one expected
from condensational growth with a mean supersaturation
\citep{hudson1995cloud,brenguier1998improvements,miles2000cloud,pawlowska2006observations, Siebert17}.
Qualitatively consistent with observations, we show that the width of droplet
size distributions broadens due to supersaturation fluctuations. 

Entrainment of dry air is not considered here. It may
lead to rapid changes of the supersaturation fluctuations
and result in an even faster broadening of the size distribution \citep{Kumar14}.
Activation of aerosols in a turbulent environment is omitted.
This may provide a more physical and realistic initial distribution of cloud droplets.
Incorporating all the cloud microphysical processes is computationally
demanding, and will have be explored in future studies.

\codedataavailability{The source code used for the simulations of this study,
the {\sc Pencil Code} \citep{axel_brandenburg_2018_2315093}, is freely available on
\url{https://github.com/pencil-code/}.
The DOI of the code is \url{http://doi.org/10.5281/zenodo.2315093}.
The last access to the code is December 16, 2018.
The DNS setup and the corresponding data \citep{li_xiang_yu_2019_2538027}
are freely available at
\url{https://doi.org/10.5281/zenodo.2538027}.}

\authorcontribution{Xiang-Yu Li developed the idea, coded the module,
performed the simulations, and wrote the manuscript. Axel Brandenburg and
Nils Haugen contributed to the development of the module and commented
on the manuscript. Gunilla Svensson contributed to the development of the idea
and commented on the manuscript.} 

\competinginterests{The authors declare that they have no conflict
of interest.}

\begin{acknowledgements}
We thank Wojtek Grabowski, Andrew Heymsfield, Gaetano Sardina, Igor Rogachevskii and Dhrubaditya Mitra
for stimulating discussions.
This work was supported through the FRINATEK grant 231444 under the
Research Council of Norway, SeRC, the Swedish Research Council grants 2012-5797 and 2013-03992,
the University of Colorado through its support of the
George Ellery Hale visiting faculty appointment,
and the grant ``Bottlenecks for particle growth in turbulent aerosols''
from the Knut and Alice Wallenberg Foundation, Dnr.\ KAW 2014.0048.
The simulations were performed using resources provided by
the Swedish National Infrastructure for Computing (SNIC)
at the Royal Institute of Technology in Stockholm and
Chalmers Centre for Computational Science and Engineering (C3SE).
This work also benefited from computer resources made available through the
Norwegian NOTUR program, under award NN9405K.
\end{acknowledgements}

\appendix

\noappendix
%

\begin{thebibliography}{47}
\providecommand{\natexlab}[1]{#1}
\providecommand{\url}[1]{{\tt #1}}
\providecommand{\urlprefix}{URL }
\expandafter\ifx\csname urlstyle\endcsname\relax
  \providecommand{\doi}[1]{https://doi.org/\discretionary{}{}{}#1}\else
  \providecommand{\doi}{https://doi.org/\discretionary{}{}{}\begingroup
  \urlstyle{rm}\Url}\fi

\bibitem[{Berry and Reinhardt(1974)}]{berry_1974}
Berry, E.~X. and Reinhardt, R.~L.: An analysis of cloud drop growth by
  collection: Part I. Double distributions, Journal of the Atmospheric
  Sciences, 31, 1814--1824, 1974.

\bibitem[{Brandenburg(2018)}]{axel_brandenburg_2018_2315093}
Brandenburg, A.: Pencil Code, \doi{10.5281/zenodo.2315093}, 2018.

\bibitem[{Brenguier et~al.(1998)Brenguier, Bourrianne, Coelho, Isbert, Peytavi,
  Trevarin, and Weschler}]{brenguier1998improvements}
Brenguier, J.-L., Bourrianne, T., Coelho, A.~A., Isbert, J., Peytavi, R.,
  Trevarin, D., and Weschler, P.: Improvements of droplet size distribution
  measurements with the Fast-FSSP (Forward Scattering Spectrometer Probe),
  Journal of Atmospheric and Oceanic Technology, 15, 1077--1090, 1998.

\bibitem[{Chandrakar et~al.(2016)Chandrakar, Cantrell, Chang, Ciochetto,
  Niedermeier, Ovchinnikov, Shaw, and Yang}]{Chandrakar16}
Chandrakar, K.~K., Cantrell, W., Chang, K., Ciochetto, D., Niedermeier, D.,
  Ovchinnikov, M., Shaw, R.~A., and Yang, F.: Aerosol indirect effect from
  turbulence-induced broadening of cloud-droplet size distributions,
  Proceedings of the National Academy of Sciences, 113, 14\,243--14\,248, 2016.

\bibitem[{Chen et~al.(2018)Chen, Yau, and Bartello}]{chen2018turbulence}
Chen, S., Yau, M., and Bartello, P.: Turbulence effects of collision efficiency
  and broadening of droplet size distribution in cumulus clouds, J. Atmosph.
  Sci., 75, 203--217, 2018.

\bibitem[{Cooper(1989)}]{cooper1989effects}
Cooper, W.~A.: Effects of variable droplet growth histories on droplet size
  distributions. Part I: Theory, Journal of the Atmospheric Sciences, 46,
  1301--1311, 1989.

\bibitem[{Desai et~al.(2018)Desai, Chandrakar, Chang, Cantrell, and
  Shaw}]{desai2018influence}
Desai, N., Chandrakar, K., Chang, K., Cantrell, W., and Shaw, R.: Influence of
  Microphysical Variability on Stochastic Condensation in a Turbulent
  Laboratory Cloud, Journal of the Atmospheric Sciences, 75, 189--201, 2018.

\bibitem[{Devenish et~al.(2012)Devenish, Bartello, Brenguier, Collins,
  Grabowski, IJzermans, Malinowski, Reeks, Vassilicos, Wang, and
  Z.Warhaft}]{DBB12}
Devenish, B., Bartello, P., Brenguier, J.-L., Collins, L., Grabowski, W.,
  IJzermans, R., Malinowski, S., Reeks, M., Vassilicos, J., Wang, L.-P., and
  Z.Warhaft: Droplet growth in warm turbulent clouds, Quart. J. Roy. Meteorol.
  Soc., 138, 1401--1429, 2012.

\bibitem[{G{\"o}tzfried et~al.(2017)G{\"o}tzfried, Kumar, Shaw, and
  Schumacher}]{Gotzfried17}
G{\"o}tzfried, P., Kumar, B., Shaw, R.~A., and Schumacher, J.: Droplet dynamics
  and fine-scale structure in a shearless turbulent mixing layer with phase
  changes, Journal of Fluid Mechanics, 814, 452--483, 2017.

\bibitem[{Grabowski and Abade(2017)}]{grabowski2017broadening}
Grabowski, W.~W. and Abade, G.~C.: Broadening of cloud droplet spectra through
  eddy hopping: Turbulent adiabatic parcel simulations, Journal of the
  Atmospheric Sciences, 74, 1485--1493, 2017.

\bibitem[{Grabowski and Wang(2013)}]{Grabowski_2013}
Grabowski, W.~W. and Wang, L.-P.: Growth of Cloud Droplets in a Turbulent
  Environment, Annu. Rev. Fluid Mech., 45, 293--324, 2013.

\bibitem[{{Haugen} et~al.(2004){Haugen}, {Brandenburg}, and
  {Dobler}}]{Haugen_etal_2004PhRvE}
{Haugen}, N.~E.~L., {Brandenburg}, A., and {Dobler}, W.: {Simulations of
  nonhelical hydromagnetic turbulence}, Phys. Rev. E, 70, 016308,
  \doi{10.1103/PhysRevE.70.016308}, 2004.

\bibitem[{Hudson and Svensson(1995)}]{hudson1995cloud}
Hudson, J.~G. and Svensson, G.: Cloud microphysical relationships in California
  marine stratus, Journal of Applied Meteorology, 34, 2655--2666, 1995.

\bibitem[{Johansen et~al.(2012)Johansen, Youdin, and Lithwick}]{Johansen_2012}
Johansen, A., Youdin, A.~N., and Lithwick, Y.: Adding particle collisions to
  the formation of asteroids and Kuiper belt objects via streaming
  instabilities, Astronomy \& Astrophysics, 537, A125,
  \doi{https://doi.org/10.1051/0004-6361/201117701}, 2012.

\bibitem[{Kabanov and Mazin(1970)}]{kabanov1970effect}
Kabanov, A. and Mazin, I.: The effect of turbulence on phase transition in
  clouds, Tr. TsAO, 98, 113--121, 1970.

\bibitem[{Katzwinkel et~al.(2014)Katzwinkel, Siebert, Heus, and
  Shaw}]{katzwinkel2014measurements}
Katzwinkel, J., Siebert, H., Heus, T., and Shaw, R.~A.: Measurements of
  turbulent mixing and subsiding shells in trade wind cumuli, Journal of the
  Atmospheric Sciences, 71, 2810--2822, 2014.

\bibitem[{Khvorostyanov and Curry(1999)}]{Khvorostyanov99}
Khvorostyanov, V.~I. and Curry, J.~A.: Toward the Theory of Stochastic
  Condensation in Clouds. Part I: A General Kinetic Equation, Journal of the
  Atmospheric Sciences, 56, 3985--3996,
  \doi{10.1175/1520-0469(1999)056<3985:TTTOSC>2.0.CO;2},
  \urlprefix\url{https://doi.org/10.1175/1520-0469(1999)056<3985:TTTOSC>2.0.CO;2},
  1999.

\bibitem[{Korolev(1995)}]{korolev1995influence}
Korolev, A.~V.: The influence of supersaturation fluctuations on droplet size
  spectra formation, Journal of the atmospheric sciences, 52, 3620--3634, 1995.

\bibitem[{Kr{\"u}ger et~al.(2017)Kr{\"u}ger, Haugen, and
  L{\o}v{\aa}s}]{kruger2017correlation}
Kr{\"u}ger, J., Haugen, N. E.~L., and L{\o}v{\aa}s, T.: Correlation effects
  between turbulence and the conversion rate of pulverized char particles,
  Combustion and Flame, 185, 160--172, 2017.

\bibitem[{Kumar et~al.(2014)Kumar, Schumacher, and Shaw}]{Kumar14}
Kumar, B., Schumacher, J., and Shaw, R.~A.: Lagrangian Mixing Dynamics at the
  Cloudy–Clear Air Interface, Journal of the Atmospheric Sciences, 71,
  2564--2580, \doi{10.1175/JAS-D-13-0294.1},
  \urlprefix\url{http://dx.doi.org/10.1175/JAS-D-13-0294.1}, 2014.

\bibitem[{{Lamb} and {Verlinde}(2011)}]{2011_lamb}
{Lamb}, D. and {Verlinde}, J.: {Physics and Chemistry of Clouds}, Cambridge,
  England, Cambridge Univ. Press, 2011.

\bibitem[{Lanotte et~al.(2009)Lanotte, Seminara, and Toschi}]{2009_Lanotte}
Lanotte, A.~S., Seminara, A., and Toschi, F.: Cloud Droplet Growth by
  Condensation in Homogeneous Isotropic Turbulence, Journal of the Atmospheric
  Sciences, 66, 1685--1697, \doi{10.1175/2008JAS2864.1},
  \urlprefix\url{http://dx.doi.org/10.1175/2008JAS2864.1}, 2009.

\bibitem[{Li et~al.(2017)Li, Brandenburg, Haugen, and
  Svensson}]{li2017eulerian}
Li, X.-Y., Brandenburg, A., Haugen, N. E.~L., and Svensson, G.: Eulerian and L
  agrangian approaches to multidimensional condensation and collection, J. Adv.
  Modeling Earth Systems, 9, 1116--1137, 2017.

\bibitem[{Li et~al.(2018)Li, Brandenburg, Svensson, Haugen, Mehlig, and
  Rogachevskii}]{li2017effect}
Li, X.-Y., Brandenburg, A., Svensson, G., Haugen, N. E.~L., Mehlig, B., and
  Rogachevskii, I.: Effect of Turbulence on Collisional Growth of Cloud
  Droplets, Journal of the Atmospheric Sciences, 75, 3469--3487,
  \doi{10.1175/JAS-D-18-0081.1},
  \urlprefix\url{https://doi.org/10.1175/JAS-D-18-0081.1}, 2018.

\bibitem[{Li et~al.(2019)Li, Svensson, Brandenburg, and
  Haugen}]{li_xiang_yu_2019_2538027}
Li, X.-Y., Svensson, G., Brandenburg, A., and Haugen, N. E.~L.: {Cloud droplet
  growth due to supersaturation fluctuations in stratiform clouds},
  \doi{10.5281/zenodo.2538027}, 2019.

\bibitem[{Marchioli et~al.(2008)Marchioli, Soldati, Kuerten, Arcen, Taniere,
  Goldensoph, Squires, Cargnelutti, and Portela}]{Marchioli08}
Marchioli, C., Soldati, A., Kuerten, J., Arcen, B., Taniere, A., Goldensoph,
  G., Squires, K., Cargnelutti, M., and Portela, L.: Statistics of particle
  dispersion in direct numerical simulations of wall-bounded turbulence:
  Results of an international collaborative benchmark test, Intern. J.
  Multiphase Flow, 34, 879--893, 2008.

\bibitem[{Miles et~al.(2000)Miles, Verlinde, and Clothiaux}]{miles2000cloud}
Miles, N.~L., Verlinde, J., and Clothiaux, E.~E.: Cloud droplet size
  distributions in low-level stratiform clouds, Journal of the atmospheric
  sciences, 57, 295--311, 2000.

\bibitem[{Paoli and Shariff(2009)}]{paoli2009turbulent}
Paoli, R. and Shariff, K.: Turbulent condensation of droplets: direct
  simulation and a stochastic model, Journal of the Atmospheric Sciences, 66,
  723--740, 2009.

\bibitem[{Pawlowska et~al.(2006)Pawlowska, Grabowski, and
  Brenguier}]{pawlowska2006observations}
Pawlowska, H., Grabowski, W.~W., and Brenguier, J.-L.: Observations of the
  width of cloud droplet spectra in stratocumulus, Geophysical Research
  Letters, 33, L19\,810, \doi{10.1029/2006GL026841}, 2006.

\bibitem[{Pope(2000)}]{Pope00}
Pope, S.: Turbulent Flows, Cambridge University Press, 2000.

\bibitem[{Pruppacher and Klett(2012)}]{pruppacher2012microphysics}
Pruppacher, H.~R. and Klett, J.~D.: Microphysics of Clouds and Precipitation:
  Reprinted 1980, Springer Science \& Business Media, 2012.

\bibitem[{Saffman and Turner(1956)}]{1955_Saffman}
Saffman, P.~G. and Turner, J.~S.: On the collision of drops in turbulent
  clouds, J. Fluid Mech., 1, 16--30, \doi{10.1017/S0022112056000020},
  \urlprefix\url{http://journals.cambridge.org/article_S0022112056000020},
  1956.

\bibitem[{Saito and Gotoh(2018)}]{saito2017turbulence}
Saito, I. and Gotoh, T.: Turbulence and cloud droplets in cumulus clouds, New
  Journal of Physics, 20, 023\,001, 2018.

\bibitem[{{Sardina} et~al.(2015){Sardina}, {Picano}, {Brandt}, and
  {Caballero}}]{2015_Sardina}
{Sardina}, G., {Picano}, F., {Brandt}, L., and {Caballero}, R.: {Continuous
  Growth of Droplet Size Variance due to Condensation in Turbulent Clouds},
  Phys. Rev. Lett., 115, 184501, \doi{10.1103/PhysRevLett.115.184501}, 2015.

\bibitem[{Sardina et~al.(2018)Sardina, Poulain, Brandt, and
  Caballero}]{sardina2018broadening}
Sardina, G., Poulain, S., Brandt, L., and Caballero, R.: Broadening of Cloud
  Droplet Size Spectra by Stochastic Condensation: Effects of Mean Updraft
  Velocity and CCN Activation, Journal of the Atmospheric Sciences, 75,
  451--467, 2018.

\bibitem[{Schiller and Naumann(1933)}]{Schiller33}
Schiller, L. and Naumann, A.: Fundamental calculations in gravitational
  processing, Zeitschrift Des Vereines Deutscher Ingenieure, 77, 318--320,
  1933.

\bibitem[{Sedunov(1965)}]{Sedunov65}
Sedunov, Y.~S.: Fine cloud structure and its role in the formation of the cloud
  spectrum, Atmos. Oceanic Phys., 1, 416--421, 1965.

\bibitem[{Seinfeld and Pandis(2016)}]{Sei06}
Seinfeld, J.~H. and Pandis, S.~N.: Atmospheric chemistry and physics: from air
  pollution to climate change, John Wiley \& Sons, 2016.

\bibitem[{Shaw(2003)}]{shaw_2003}
Shaw, R.~A.: Particle-turbulence interactions in atmospheric clouds, Annu. Rev.
  Fluid Mech., 35, 183--227, 2003.

\bibitem[{{Shima} et~al.(2009){Shima}, {Kusano}, {Kawano}, {Sugiyama}, and
  {Kawahara}}]{Shima09}
{Shima}, S., {Kusano}, K., {Kawano}, A., {Sugiyama}, T., and {Kawahara}, S.:
  {The super-droplet method for the numerical simulation of clouds and
  precipitation: a particle-based and probabilistic microphysics model coupled
  with a non-hydrostatic model}, Quart. J. Roy. Met. Soc., 135, 1307--1320,
  2009.

\bibitem[{Siebert and Shaw(2017)}]{Siebert17}
Siebert, H. and Shaw, R.~A.: Supersaturation fluctuations during the early
  stage of cumulus formation, Journal of the Atmospheric Sciences, 74,
  975--988, 2017.

\bibitem[{Siewert et~al.(2017)Siewert, Bec, and
  Krstulovic}]{siewert2017statistical}
Siewert, C., Bec, J., and Krstulovic, G.: Statistical steady state in turbulent
  droplet condensation, Journal of Fluid Mechanics, 810, 254--280, 2017.

\bibitem[{Srivastava(1989)}]{srivastava1989growth}
Srivastava, R.: Growth of cloud drops by condensation: A criticism of currently
  accepted theory and a new approach, Journal of the atmospheric sciences, 46,
  869--887, 1989.

\bibitem[{Vaillancourt et~al.(2001)Vaillancourt, Yau, and
  Grabowski}]{vaillancourt2001microscopic}
Vaillancourt, P., Yau, M., and Grabowski, W.~W.: Microscopic approach to cloud
  droplet growth by condensation. Part I: Model description and results without
  turbulence, Journal of the atmospheric sciences, 58, 1945--1964, 2001.

\bibitem[{Vaillancourt et~al.(2002)Vaillancourt, Yau, Bartello, and
  Grabowski}]{vaillancourt2002microscopic}
Vaillancourt, P., Yau, M., Bartello, P., and Grabowski, W.~W.: Microscopic
  approach to cloud droplet growth by condensation. Part II: Turbulence,
  clustering, and condensational growth, Journal of the atmospheric sciences,
  59, 3421--3435, 2002.

\bibitem[{{Veysey} and Goldenfeld(2007)}]{John07}
{Veysey}, II, J. and Goldenfeld, N.: Simple viscous flows: From boundary layers
  to the renormalization group, Rev. Modern Phys., 79, 883--927, 2007.

\bibitem[{Yau and Rogers(1996)}]{yau1996short}
Yau, M.~K. and Rogers, R.: A short course in cloud physics, Elsevier, 1996.

\end{thebibliography}

%
\end{document}